\documentclass{article}
\usepackage{spconfa4,amsmath,graphicx,url,times, amssymb, latexsym}
\usepackage{color}

\title{Maximum Likelihood Estimation of the   Direction of Sound In A Reverberant  Noisy Environment}

\name{Mohamed F. Mansour}
\address{Amazon Inc., USA}

\begin{document}

\ninept
\maketitle

\begin{sloppy}

\begin{abstract}
  We describe a new method for estimating the direction of sound in a reverberant environment from basic  principles of sound propagation. The method utilizes SNR-adaptive features from time-delay and energy of the directional components after acoustic wave decomposition of the observed sound field to estimate the line-of-sight direction under noisy and reverberant conditions. The effectiveness of the approach is established with measured data of different microphone array configurations under various  usage scenarios.
\end{abstract}


\vspace{-0.2cm}
\section{Introduction}
%\vspace{-0.1cm}
 Computing the direction of arrival (DoA) of a sound source is a classical estimation  problem that is essential for sound source localization. It has many applications in robotics and speech communication systems \cite{argentieri2015survey, rascon2017localization}, and has become increasingly important with the proliferation of voice-controlled smart systems \cite{chhetriAFE2018Eusipco}.   Many techniques have been proposed to address the problem including, beamforming \cite{dmochowski2007generalized, daniel2020time}, subspace methods, e.g., MUSIC and ESPRIT \cite{argentieri2007broadband, hogg2021polynomial}, time delay methods, e.g., GCC-PHAT and SRP-PHAT \cite{knapp1976generalized, dibiase2001robust}, and more recently DNN methods \cite{yalta2017sound, grumiaux2022survey}. In the absence of strong reverberation/interference, existing techniques generally provide satisfactory results, and this has been studied extensively in the literature. However, a commercial-grade embedded system for sound localization, with constraints on computation/latency/memory, requires consistent localization performance  under adverse reverberation and noise conditions, and this is the subject of this work.
 
The fundamental problem in computing the DoA of a sound source in a reverberant and noisy environment  is to distinguish the line-of-sight component of the target sound source from all interfering directional components in the presence of incoherent sensor noise. These interfering directional components include acoustic reflections of the target sound source, as well as all directional components of coherent noise interference. Hence, all solutions to the DoA problem aim at finding a proper characterization of the line-of-sight component based on either a physical model or a data-driven model.
Signal processing solutions to the problem, e.g., SRP-PHAT and MUSIC algorithms deploy a channel model for acoustic propagation and source/noise statistics, where it is generally assumed that the direct path component is on average stronger than acoustic reflections across a range of frequencies of interest. The direct path component is computed implicitly using inter-microphone information, e.g., generalized cross-correlation. These approaches frequently fail to accurately capture common cases, e.g., when there is a strong room reverberation and the microphone array is placed at a corner far from the source. This problem is more apparent with small microphone arrays, e.g., $ \le 4$ microphones, where due to the coarse sensing resolution, the line-of-sight component might be perceived as weaker than some reflections. Moreover, the problem gets more complicated in the presence of coherent interference with speech-like content,  e.g., from TV. To resolve some of these issues, data-driven approaches that deploy variations of deep neural networks (DNN) were introduced in recent years, with the assumption that training data captures all relevant use cases. These approaches showed improvement (sometimes significant) over classical approaches on the test datasets \cite{grumiaux2022survey} especially under noisy conditions. However, as noted in \cite{wu2021sslide} that all these solutions  can only work well when distance between source and microphone array is small, which is a limitation for commercial adoption. Further, this approach is not scalable to accommodate different microphone array geometries, as the training data is dependent on the microphone array and it should capture a huge number of cases  that cover all usage scenarios at different kinds of rooms, noise, and sound stimuli. Unlike training of speech models for ASR, synthetic models, e.g., image source method \cite{imagemethod}, cannot replace data collections because the learning objective is the model itself and they are parameterized by a relatively small number of parameters that can be learned by the DNN model. 
%Moreover, these approaches cannot accommodate failed/degraded microphones, and the performance can degrade with different sound stimuli if not properly captured in the training dataset.

The work presented in this paper provides a new methodology for computing the sound direction that is based on directional decomposition of microphone array observations. The microphone array observations are mapped to directional component via acoustic wave decomposition, and these directional components are processed to compute the sound direction. The multidimensional representation of the spatial signal with directional components provides an intuitive characterization of the line-of-sight component of the sound source based on principles of acoustic propagation, which was not explored in earlier works. It utilizes a generalized acoustic propagation model that accommodates total acoustic pressure due to scattering at the mounting surface. This physical characterization is utilized to construct a statistical framework to derive the maximum-likelihood estimator of the direction of arrival.  The mapping to directional components is empowered by the work in \cite{chhetri2019acoustic, icassp_awd}, where a method for generalized acoustic wave decomposition of a microphone array of arbitrary geometry was described.  It does not require a special microphone array geometry as in related localization work with spherical harmonics \cite{jarrett20103d}.  The proposed system is suited for embedded implementation and it is scalable to accommodate different microphone array geometries with minimal tuning effort.  The discussion in this work is limited single-source localization. It is shown in section \ref{sec:results} that the proposed algorithm outperforms existing baseline solutions in mitigating large localization errors when evaluated on a large corpus of real data under diverse room conditions and different microphone array size and geometry.  

\vspace{-2mm}
 \section{Problem Definition}\label{sec:prob_def}
\vspace{-1mm}
 
The underlying physical model of the estimation problem is the generalized Acoustic Wave Decomposition (AWD) as described in \cite{chhetri2019acoustic, icassp_awd}, where the observed sound field, ${\mathbf{p}}(\omega; t)$, at the microphone array is expressed as
\begin{equation}
\vspace{-2mm}
{\mathbf{p}}(\omega; t) = \sum_{l}  \alpha_l(\omega;t) \ \boldsymbol{\psi}(\omega; \theta_l(t), \phi_l(t)) \label{eq:awd}
\vspace{-1mm}
\end{equation}
where $\theta_l$ and $\phi_l$ denote respectively the elevation and azimuth (in polar coordinates) of the direction of propagation of the $l$-th acoustic wave, and $\boldsymbol{\psi}(\omega; \theta,\phi)$ denotes the total acoustic pressure at the microphone array when a free-field acoustic plane wave with direction $(\theta, \phi)$  impinges on the device. The total acoustic pressure is the superposition of the incident free-field plane wave and the scattered component at the device surface. At each $\omega$, $\boldsymbol{\psi}(\omega; .)$ is a vector whose length equals the number of microphones in the microphone array. The ensemble of all vectors that span the three-dimensional space at all $\omega$ defines the \emph{acoustic dictionary} of the device, and it is computed offline with standard acoustic simulation techniques \cite{chhetri2019acoustic, icassp_awd}. Note that, even though the elevation is not reported in the direction of sound, it is important to include it in the signal model, as acoustic waves with the same azimuth but different elevation might have different impact at the microphone array when surface scattering is accommodated.

This model generalizes the free-field plane wave decomposition to accommodate scattering at the device surface which is modeled as a hard boundary.  
This scattering component partially resolves spatial aliasing due to phase ambiguity at high frequencies.
Each directional component of the acoustic wave expansion in \eqref{eq:awd} at frame $t$ is characterized by its direction $\left(\theta_l(t), \phi_l(t)\right)$ (which are frequency-independent), and the corresponding complex-valued weight $\alpha_l(\omega;t)$.

The objective of sound direction estimation  is to compute the azimuth angle $\hat{\phi}$ that corresponds to the line-of-sight direction of the sound source, given the observed sound field $\{\mathbf{p}(\omega; t)\}_{\omega, t}$ at successive time frames that span the duration of the source signal. In this work, we assume only a single sound source, though background coherent or incoherent noise can be present.

In the absence of other sound sources, the line-of-sight component of a sound source is usually contaminated by other directional components due to acoustic reflections at nearby surfaces, as well as incoherent noise at the microphone array. Nevertheless, the line-of-sight component has two distinct features:
\begin{enumerate}
\item The energy of the line-of-sight component of a sound source is higher than the energy of any of its individual reflections.
\item The line-of-sight component of a sound source arrives first at the microphone array before any other reflection of the same sound source.
\end{enumerate}
The following section describes a statistical framework that utilizes these two features to design a maximum-likelihood estimator of the sound direction
%These two features are used to design a maximum-likelihood estimator of the sound direction, by combining two likelihood functions of time-delay and energy for each directional component. 
\vspace{-2mm}
 \section{Maximum-Likelihood Estimation}\label{sec:algo}
 
 \subsection{Estimation Procedure}
The estimation procedure  exploits the true direction features as described in the previous section to compute the maximum-likelihood estimate of the user direction. It computes from microphone array observations two likelihood functions for the time delay and the signal energy; then applies late fusion to compute the total likelihood at each time frame. The two likelihood functions are computed from the directional components in \eqref{eq:awd} at each time frame. Finally, the total likelihood values at different frames are smoothed over the duration of the sound signal to produce the aggregate likelihood function that is used to find the maximum-likelihood estimate. Hence, the estimation flow is as follows:
\begin{enumerate}
\item At each time step $t$, process the observed sound field, $\{\mathbf{p}(\omega; t)\}_\omega$, as follows
\begin{enumerate}
\item Compute the acoustic wave decomposition at all $\omega$  in \eqref{eq:awd} using the multistage solver as  in \cite{icassp_awd}.
\item Compute the time-delay likelihood function of each directional component  (as described in section \ref{sec:td-like}).
\item Compute the energy-based likelihood function of each directional component (as described in section \ref{sec:pwr-like}).
\item Combine the two likelihood functions on a one-dimensional grid of possible azimuth angles (as described in section \ref{sec:local-like}).
\end{enumerate}
\item Compute the aggregate likelihood of each angle candidate over the whole signal period, and choose the angle that corresponds to the maximum-likelihood value as the sound direction estimate (as outlined in section \ref{sec:local-like}).
\end{enumerate} 

\subsection{Delay-based Likelihood \label{sec:td-like}}
Assume that a source  signal, $X(\omega)$,  experiences multiple reflections in the acoustic path towards a microphone array. Denote the $l$-th reflection at a receiving microphone by $X_l(\omega)$, which can be expressed as
\begin{equation}
X_l(\omega) \approx  e^{-j\omega\tau_l} \  \delta_l(\omega) \ X(\omega) \label{eq:sigmodel}
\end{equation}
where $\tau_l > 0$ is the corresponding delay, and $\delta_l$ is a real-valued propagation loss. Note that, $X_l(\omega)$ refers to the $l$-th AWD component in \eqref{eq:awd} due to sound source $X(\omega)$. To eliminate the nuisance parameters $\delta_l$ and $X(\omega)$, we introduce the auxiliary parameter
\begin{eqnarray}
Q_{lk} (\omega) &\triangleq& \frac{X_l(\omega) / \|X_l(\omega)\|}{X_k(\omega) /  \|X_k(\omega)\|} \nonumber \\
	                  &\approx& e^{j\omega (\tau_k-\tau_l)} %\ \frac{\delta_{l}(\omega)}{\delta_{k}(\omega)}
\end{eqnarray}
%which could be further simplified as 
%\begin{eqnarray}
%\overline Q_{lk} (\omega) &\triangleq& \frac{Q_{lk} (\omega) }{\|Q_{lk} (\omega) \|}   \nonumber \\
%                                 &\approx& e^{j\omega (\tau_k-\tau_l)}
%\end{eqnarray}
This parameter is utilized to find the time delay between the two components. However, it is susceptible to phase wrapping at large $\omega$, and  one extra step is needed to mitigate its impact. Define for a frequency shift $\Delta$
\begin{eqnarray}
R_{lk}(\omega) &\triangleq& \frac{ Q_{lk} (\omega+\Delta)}{ Q_{lk} (\omega)} \nonumber \\
                        &\approx& e^{j\Delta (\tau_k-\tau_l)}
\end{eqnarray}
which eliminates the dependence on $\omega$, and if $\Delta$ is chosen small enough, then phase wrapping is eliminated. Then, the estimated delay between components $l$ and $k$, $\bar\rho_{lk} \triangleq \tau_k -\tau_l$, is computed as
\begin{equation}
\bar\rho_{lk} =  \frac{1}{\sum_{\omega \in \Omega} W(\omega)} \ \ \sum_{\omega \in \Omega} W(\omega) \ \frac{\angle R_{lk}(\omega)}{\Delta}     \label{eq:rholk}
\end{equation}
where $\angle R$ denotes the angle of $R$, and $W(\omega)$ is a sigmoid weighting function that depends on SNR at $\omega$. Note that, the procedure does not require computing the inverse FFT as in common generalized cross-correlation schemes \cite{brutti2008comparison}, which significantly reduces the overall complexity.
If $\bar\rho_{lk} > 0$, then the $k$-th reflection is delayed from the $l$-th reflection and vice versa. Hence, the probability that the $k$-th component is delayed from the $l$-th component is $P(\rho_{lk} > 0)$ (where $\rho_{lk}$ is the true value of $\tau_k - \tau_l$). If $\rho_{lk} \sim \mathcal{N}(\bar\rho_{lk}, \sigma^2)$, then
\begin{equation}
P(\tau_k > \tau_l) \equiv P(\rho_{lk} > 0) \approx \frac{1}{2}\ \text{erfc} \left(\frac{-\bar\rho_{lk}}{\sigma\sqrt{2}}\right) \label{eq:p_rho}
\end{equation}
where $\text{erfc}(.)$ is the complementary error function. Note that, $P(\tau_k < \tau_l) = 1-P(\tau_k > \tau_l) $, hence, $\bar\rho_{lk}$ is computed once for each pair of components.
%\subsubsection{Time-Delay Likelihood Computation}

The acoustic reflections $\{X_l(\omega)\}$ are approximated by $\{\alpha_l(\omega; t)\}_\omega$ in \eqref{eq:awd}. Denote the probability that the $l$-th component is the first to arrive at the microphone array by $\beta_l$, which can be expressed as
\begin{eqnarray}
 \beta_l &\triangleq& P(\tau_l < \min\{\tau_k\}_{k\neq l}) \nonumber \\
                &=& \prod_{k\neq l} P(\rho_{lk} > 0) 
 \end{eqnarray}
 which, using \eqref{eq:p_rho}, can be expressed in the log-domain as 
\begin{equation}
\bar\beta_{l} \approx \sum_{k \neq l} \log \left( \text{erfc} \left(\frac{-\bar\rho_{lk}}{\sigma\sqrt{2}}\right)  \right) \label{eq:beta}
\end{equation}  
This is a good approximation of the time-delay likelihood function as long as the AWD components follow the signal model in \eqref{eq:sigmodel}. A simple test to validate this assumption is to compute the pair-wise correlation coefficient between components, and run the computation only if it is above a predetermined threshold.

\subsection{Energy-based Likelihood \label{sec:pwr-like}}
The true energy of the line-of-sight component is theoretically higher than the energy of each individual reflection. However, due to the finite number of microphones, the true directional component might be diluted in the AWD computation. Nevertheless, the line-of-sight energy is usually among the highest energy components. The energy of each component is computed as
\begin{equation}
E_l = \sum_{\omega} \| \alpha_l(\omega; t)\|^2 \ W(\omega)
\end{equation}
where $W(\omega)$ is a weighting function as in \eqref{eq:rholk}. An AWD component is a candidate to be the line-of-sight component if $E_l > \nu \ \max\{E_k\}$, where $\nu$ is a predetermined threshold. All directional components above the energy threshold are considered equally likely to be the line-of-sight component. Hence, if the number of AWD components that satisfy this condition is $M$, then the energy-based likelihood is computed as
\begin{equation}
\gamma_l = \left\{ \begin{array}{llr}
		   -\log M & \mbox{if}&   E_l > \nu \ \max\{E_k\} \\
		   \varepsilon & \mbox{otherwise}& \end{array} \right.   \label{eq:gamma}
\end{equation}
where $\varepsilon \ll -\log M$ corresponds to a small probability value to account for measurement/computation errors.
  
\subsection{Aggregate Likelihood \label{sec:local-like}} 
At each time frame, the log-likelihoods $\bar\beta_l$ and $\gamma_l$ are computed for the $l$-th AWD component as in \eqref{eq:beta} and \eqref{eq:gamma} respectively. This corresponds to azimuth angle $\phi_l$ of the corresponding entry of the device dictionary, and the total likelihood at $\phi_l$ is the sum of the two likelihoods.
Due to the finite dictionary size and the finite precision of the computation, the true angle of the $l$-th component can be an angle adjacent to $\phi_l$. If we assume normal distribution (with variance $\kappa$) of the true azimuth angle around $\phi_l$, then the likelihood for azimuth angles adjacent to $\phi_l$ is approximated as
\begin{equation}
\chi(\phi; t) = \max\left(\chi(\phi; t)\ ; \ \beta_l + \gamma_l- \frac{(\phi-\phi_l)^2}{2\kappa}\right).  \label{eq:diff-like}
\end{equation}
and the likelihood function, $\chi(.)$, of all azimuth angles is updated according to \eqref{eq:diff-like} with every new AWD component. Note that, the azimuth likelihood is smoothed with the azimuth angle of each AWD component, $\phi_l$, whereas  the elevation component, $\theta_l$, is treated as nuisance parameter that is averaged out. If joint estimation of azimuth and elevation is required, then a two-dimensional likelihood function $\chi(\theta,\phi; t)$ is utilized rather than the one-dimensional likelihood in \eqref{eq:diff-like}.

The final step is to compute the maximum-likelihood estimate of the azimuth angle by aggregating the local likelihood values in \eqref{eq:diff-like} over the duration of the sound event. The final likelihood aggregates the local likelihood at different time frames after proper weighting by the total SNR at each frame. 
\begin{equation}
\bar{\chi}(\phi) = \sum_t \chi(\phi; t) \ \eta(t)   \label{eq:agg-like}
\end{equation}
where $\eta(t)$ is a weighting sigmoid function that is proportional to the total SNR at frame $t$. This temporal weighting is necessary to mitigate errors in the sound event time boundaries. 
The maximum-likelihood estimate of the angle of arrival is computed from \eqref{eq:agg-like} as 
\begin{equation}
\hat{\phi} = \text{argmax } \ \bar{\chi}(\phi)
\end{equation}

\subsection{Discussion}
The signal measurement model for the maximum-likelihood estimation is the general physical model in \eqref{eq:awd}, and the properties of the line-of-sight component as described in section \ref{sec:prob_def}.
 %The estimation algorithm as described in this section exploits the general physical model in \eqref{eq:awd}, and the properties of the line-of-sight component as described in section \ref{sec:prob_def}. 
 The formulation of a statistical model, in the form of the combined likelihood function in \eqref{eq:diff-like} from this physical model is the key contribution of this work. 
 
 The aggregate likelihood function combines time-delay likelihood and energy likelihood to capture the physical properties of the line-of-sight component. These likelihood computations utilize the directional components from the acoustic wave decomposition in \eqref{eq:awd} as described in \cite{icassp_awd}. The reverberation impact is mitigated by incorporating the time-delay component, while the noise impact is mitigated by incorporated SNR-dependent weighting. %The memory and computational complexity of the algorithm is scalable as described in \cite{icassp_awd}.
 The algorithm is fundamentally different from existing model-based algorithms in few aspects:
 \begin{itemize}
 \item Incorporating both time-delay and energy to compute the direction of sound. 
\item Incorporating magnitude component in steering vectors to reduce spatial aliasing. 
\item Utilizing sparse techniques to find relevant directions in 3D space, rather than deploying exhaustive beamforming that unavoidably has spatial leakage from adjacent directions, 
\item It scales properly with minimal tuning to other microphone array geometries through the acoustic dictionary. 
\end{itemize}

 \section{Experimental Evaluation} \label{sec:results}
 The proposed algorithm is evaluated using two different microphone arrays. The first microphone array is a circular array with $8$ microphones mounted atop a cylindrical surface. The second microphone array is a star-shaped 3D array with $4$ microphones that are mounted on an LCD screen. The geometry and mounting surface of the two arrays are quite different to illustrate the generality of the proposed method. The test dataset has approximately $55$k utterances recorded in different rooms at different SNR levels. The dataset covers all angles around the microphone array, and covers possible placements of the microphone array inside a room, i.e.,  center, wall, and corner. For each microphone array, the device acoustic dictionary is computed offline as described in section \ref{sec:prob_def}. The noise was recorded at the microphone array separately and added to clean speech for evaluation. It covers a wide set of household noises, e.g., fan, vacuum, TV, microwave, ... etc. 
  
 Fig. \ref{fig:mae} shows the average performance of the proposed algorithm at different SNR values. In both microphone array cases, the mean absolute error is around $6^\circ$  at high SNR and it degrades gracefully with lower SNR. The 8-mic configuration provides $5$ to $9$ dB advantage over the 4-mic configuration depending on the operating SNR. Note that, the proposed algorithm does not explicitly deploy a denoising mechanism. Rather, it deploys a noise mitigation mechanism through the SNR-dependent weighting as discussed in section \ref{sec:algo}. The performance at low SNR can be improved with a denoising procedure prior to estimation but this is outside the scope of this work. 
 
 \begin{figure}[h]
	%\vspace{-2mm} 
	\centering
	\includegraphics[width=7.0cm, height=5.0cm, ,trim=10mm 3mm 10mm 10mm,clip]{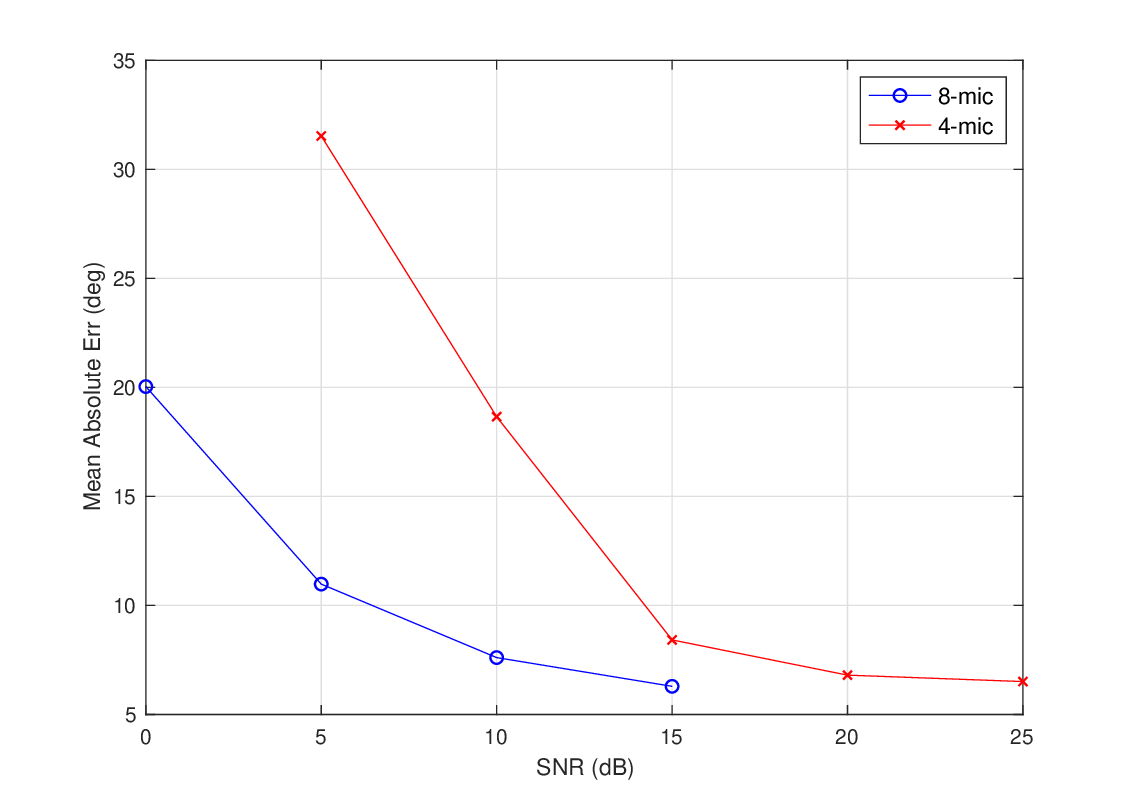}
	\caption{Mean Absolute Error in degrees of the proposed algorithm for microphone arrays of size $8$ and $4$}
	\label{fig:mae}
	%\vspace{-5mm} 
\end{figure}

In Fig. \ref{fig:cdf}, the cumulative density function (CDF) of the absolute error for the proposed algorithm is shown. It is compared to the CDF of the SRP-PHAT and state-of-the-art DNN solution. In both cases, the compared algorithms are fully tuned to the respective microphone array by subject matter experts for the respective hardware. The DNN solution utilizes an enhanced implementation of CRNN-SSL algorithm in \cite{adavanne2018sound} with few architecture changes to match state of the art performance. It is trained with data from the device with $4$-mic microphone array at numerous room configurations with a combination of synthetic and real-data. The SRP-PHAT solution utilizes heuristics to increase robustness to strong reflections and interfering noise.  As shown in the figure, the proposed algorithm provides improvement in both cases especially for the high-error case, which corresponds to low-SNR cases and cases with strong reverberation. For example, both the $90$-percentile and the $95$-percentile errors are reduced by more than $50\%$ as compared to SRP-PHAT as illustrated in Fig. \ref{fig:cdf}a. Hence, the proposed algorithm is more effective in mitigating large estimation errors, which usually have big negative impact on the user experience.

 \begin{figure}[h]
	%\vspace{-2mm} 
	%\centering
	\includegraphics[width=8.5cm, height=4.5cm, ,trim=29mm 0mm 27mm 0mm,clip]{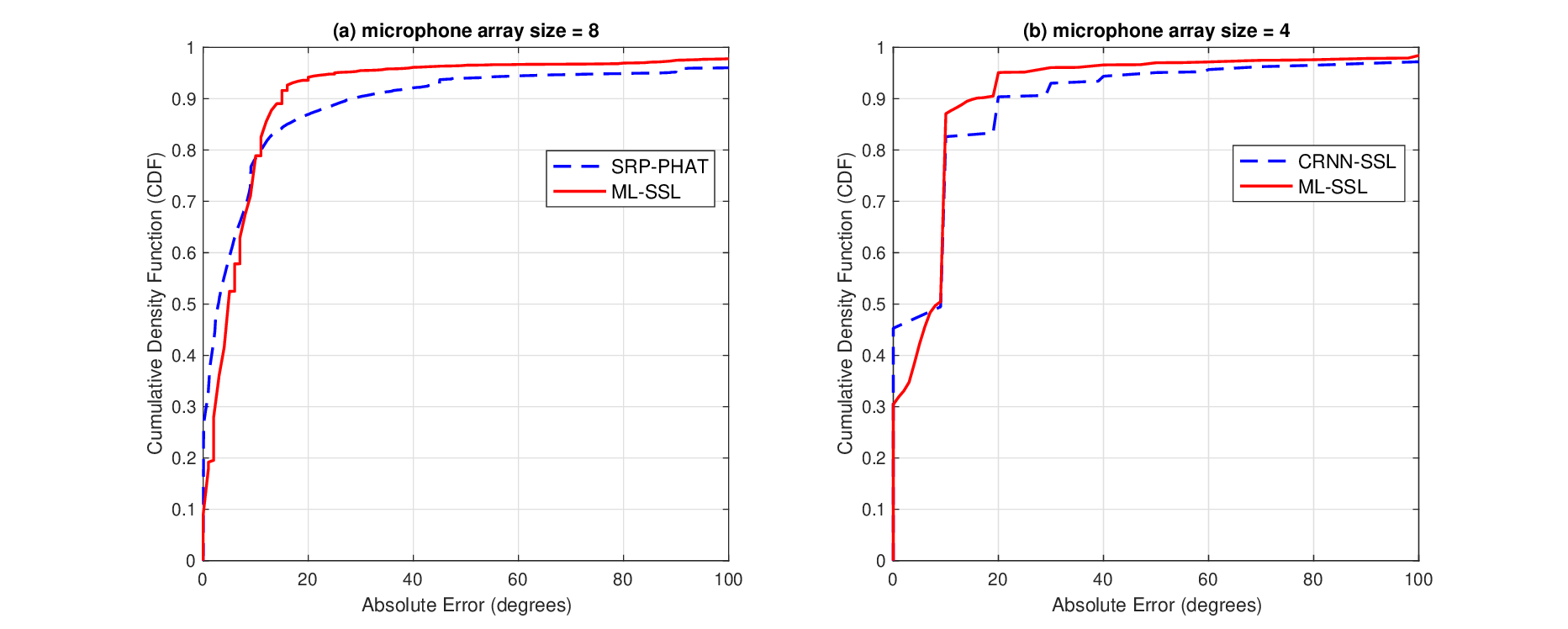}
	\caption{Cumulative density function of the absolute error for proposed algorithm vs (a.) SRP-PHAT with 8-mic, (b) CRNN-SSL with 4-mic}
	\label{fig:cdf}
	%\vspace{-5mm} 
\end{figure}

\section{Conclusion}
The proposed algorithm  addresses the two fundamental problems in computing sound source direction, namely reverberation and noise interference. It is founded on a rigorous and general physical model for sound propagation, which is mapped to a statistical model that is solved by standard estimation techniques to compute the maximum-likelihood estimation. The proposed algorithm is shown to outperform existing solutions in the literature when evaluated with a large dataset of real data.
 
Further, the proposed algorithm has two practically important advantages over prior art:
\begin{enumerate}
\item It is agnostic to the geometry of the microphone array and mounting surface because the input to the estimation procedure is the directional components after wave decomposition rather than microphone array observations. The device dependent part is captured in the device acoustic dictionary, which does not contribute to the algorithm hyper parameters. This enhances scalability and reduces migration effort to new hardware designs.
\item It generalizes the acoustic propagation model to accommodate scattering at the device surface. This scattering is viewed as distortion if free-field propagation model is utilized, whereas it is leveraged in the proposed system to enhance estimation. The incorporation of the magnitude components, due to scattering, in addition to phase components enhances robustness to spatial aliasing.
\end{enumerate}

 The proposed algorithm does not deploy a noise enhancement procedure prior to estimation. A multichannel signal enhancement system can improve the performance at low SNR if it preserves the coherence between microphones, and this is a subject of future work. Future work also includes utilizing of directional components for joint source localization and separation.

% -------------------------------------------------------------------------
% Either list references using the bibliography style file IEEEtran.bst
\bibliographystyle{IEEEtran}
%\bibliography{refs23}
%\bibliographystyle{IEEEbib}
\bibliography{refs}

\begin{thebibliography}{10}
\providecommand{\url}[1]{#1}
\def\UrlFont{\rmfamily}
\providecommand{\newblock}{\relax}
\providecommand{\bibinfo}[2]{#2}
\providecommand\BIBentrySTDinterwordspacing{\spaceskip=0pt\relax}
\providecommand\BIBentryALTinterwordstretchfactor{4}
\providecommand\BIBentryALTinterwordspacing{\spaceskip=\fontdimen2\font plus
\BIBentryALTinterwordstretchfactor\fontdimen3\font minus
  \fontdimen4\font\relax}
\providecommand\BIBforeignlanguage[2]{{%
\expandafter\ifx\csname l@#1\endcsname\relax
\typeout{** WARNING: IEEEtran.bst: No hyphenation pattern has been}%
\typeout{** loaded for the language `#1'. Using the pattern for}%
\typeout{** the default language instead.}%
\else
\language=\csname l@#1\endcsname
\fi
#2}}

\bibitem{argentieri2015survey}
S.~Argentieri, P.~Danes, and P.~Sou{\`e}res, ``A survey on sound source
  localization in robotics: From binaural to array processing methods,''
  \emph{Computer Speech \& Language}, vol.~34, no.~1, pp. 87--112, 2015.

\bibitem{rascon2017localization}
C.~Rascon and I.~Meza, ``Localization of sound sources in robotics: A review,''
  \emph{Robotics and Autonomous Systems}, vol.~96, pp. 184--210, 2017.

\bibitem{chhetriAFE2018Eusipco}
A.~Chhetri, P.~Hilmes, T.~Kristjansson, W.~Chu, M.~Mansour, X.~Li, and
  X.~Zhang, ``Multichannel {A}udio {F}ront-{E}nd for {F}ar-{F}ield {A}utomatic
  {S}peech {R}ecognition,'' in \emph{2018 European Signal Processing Conference
  (EUSIPCO)}, 2018, pp. 1527--1531.

\bibitem{dmochowski2007generalized}
J.~Dmochowski, J.~Benesty, and S.~Affes, ``A generalized steered response power
  method for computationally viable source localization,'' \emph{IEEE
  Transactions on Audio, Speech, and Language Processing}, vol.~15, no.~8, pp.
  2510--2526, 2007.

\bibitem{daniel2020time}
J.~Daniel and S.~Kiti{\'c}, ``Time domain velocity vector for retracing the
  multipath propagation,'' in \emph{ICASSP 2020-2020 IEEE International
  Conference on Acoustics, Speech and Signal Processing (ICASSP)}.\hskip 1em
  plus 0.5em minus 0.4em\relax IEEE, 2020, pp. 421--425.

\bibitem{argentieri2007broadband}
S.~Argentieri and P.~Danes, ``Broadband variations of the music high-resolution
  method for sound source localization in robotics,'' in \emph{2007 IEEE/RSJ
  International Conference on Intelligent Robots and Systems}.\hskip 1em plus
  0.5em minus 0.4em\relax IEEE, 2007, pp. 2009--2014.

\bibitem{hogg2021polynomial}
A.~Hogg, V.~Neo, S.~Weiss, C.~Evers, and P.~Naylor, ``A polynomial eigenvalue
  decomposition music approach for broadband sound source localization,'' in
  \emph{2021 IEEE Workshop on Applications of Signal Processing to Audio and
  Acoustics (WASPAA)}.\hskip 1em plus 0.5em minus 0.4em\relax IEEE, 2021, pp.
  326--330.

\bibitem{knapp1976generalized}
C.~Knapp and G.~Carter, ``The generalized correlation method for estimation of
  time delay,'' \emph{IEEE transactions on acoustics, speech, and signal
  processing}, vol.~24, no.~4, pp. 320--327, 1976.

\bibitem{dibiase2001robust}
J.~DiBiase, H.~Silverman, and M.~Brandstein, ``Robust localization in
  reverberant rooms,'' in \emph{Microphone arrays}.\hskip 1em plus 0.5em minus
  0.4em\relax Springer, 2001, pp. 157--180.

\bibitem{yalta2017sound}
N.~Yalta, K.~Nakadai, and T.~Ogata, ``Sound source localization using deep
  learning models,'' \emph{Journal of Robotics and Mechatronics}, vol.~29,
  no.~1, pp. 37--48, 2017.

\bibitem{grumiaux2022survey}
P.~Grumiaux, S.~Kiti{\'c}, L.~Girin, and A.~Gu{\'e}rin, ``A survey of sound
  source localization with deep learning methods,'' \emph{The Journal of the
  Acoustical Society of America}, vol. 152, no.~1, pp. 107--151, 2022.

\bibitem{wu2021sslide}
Y.~Wu, R.~Ayyalasomayajula, M.~Bianco, D.~Bharadia, and P.~Gerstoft, ``Sslide:
  Sound source localization for indoors based on deep learning,'' in
  \emph{ICASSP 2021-2021 IEEE International Conference on Acoustics, Speech and
  Signal Processing (ICASSP)}.\hskip 1em plus 0.5em minus 0.4em\relax IEEE,
  2021, pp. 4680--4684.

\bibitem{imagemethod}
J.~Allen and D.~Berkley, ``Image method for efficiently simulating small-room
  acoustics,'' \emph{The Journal of the Acoustical Society of America},
  vol.~65, no.~4, pp. 943--950, 1979.

\bibitem{chhetri2019acoustic}
A.~Chhetri, M.~Mansour, W.~Kim, and G.~Pan, ``On acoustic modeling for
  broadband beamforming,'' in \emph{27th European Signal Processing Conference
  (EUSIPCO)}, 2019, pp. 1--5.

\bibitem{icassp_awd}
M.~Mansour, ``Sparse recovery of acoustic waves,'' in \emph{ICASSP 2022-2022
  IEEE International Conference on Acoustics, Speech and Signal Processing
  (ICASSP)}.\hskip 1em plus 0.5em minus 0.4em\relax IEEE, 2022, pp. 5418--5422.

\bibitem{jarrett20103d}
D.~Jarrett, E.~Habets, and P.~Naylor, ``3d source localization in the spherical
  harmonic domain using a pseudointensity vector,'' in \emph{2010 18th European
  Signal Processing Conference}.\hskip 1em plus 0.5em minus 0.4em\relax IEEE,
  2010, pp. 442--446.

\bibitem{brutti2008comparison}
A.~Brutti, M.~Omologo, and P.~Svaizer, ``Comparison between different sound
  source localization techniques based on a real data collection,'' in
  \emph{2008 hands-free speech communication and microphone arrays}.\hskip 1em
  plus 0.5em minus 0.4em\relax IEEE, 2008, pp. 69--72.

\bibitem{adavanne2018sound}
S.~Adavanne, A.~Politis, J.~Nikunen, and T.~Virtanen, ``Sound event
  localization and detection of overlapping sources using convolutional
  recurrent neural networks,'' \emph{IEEE Journal of Selected Topics in Signal
  Processing}, vol.~13, no.~1, pp. 34--48, 2018.

\end{thebibliography}

\end{sloppy}
\end{document}